\documentclass[preprint2]{aastex}

\shorttitle{Pristine CNO abundances from MC B stars II.}
\shortauthors{Korn et al.}

\begin{document}


\title{Pristine CNO abundances from Magellanic Cloud B stars\\
II.~Fast rotators in the LMC cluster NGC\,2004\,\footnote{Based on observations carried out at the European Southern Observatory (ESO), Paranal, Chile, under programme ID 66.D-0214(A).}}


\author{A. J. Korn}
\affil{Uppsala Astronomical Observatory (UAO), Box 515, 75120 Uppsala, Sweden}
\email{akorn@astro.uu.se}

\author{M. F. Nieva\footnote{current address: Dr.~Remeis-Sternwarte Bamberg, Sternwartstr.7, 96049 Bamberg, Germany}, S. Daflon and K. Cunha\footnote{current address: National Optical Astronomy Observatory, P.O. Box 26732, Tucson, AZ 85726, USA}}
\affil{Observat\'{o}rio Nacional, Rua General Jos\'{e} Cristino 77, CEP 20921-400, Rio de Janeiro, Brazil}
\email{fnieva@on.br, daflon@on.br, katia@on.br}



\begin{abstract}
We present spectroscopic abundance analyses of three main-sequence B stars in the young Large Magellanic Cloud cluster NGC\,2004. All three targets have projected rotational velocities around 130\,km\,s$^{-1}$. Techniques are presented that allow the derivation of stellar parameters and chemical abundances in spite of these high $v\,\sin i$ values. Together with previous analyses of stars in this cluster, we find no evidence among the main-sequence stars for effects due to rotational mixing up to $v\,\sin i$\,$\approx$\,130\,km\,s$^{-1}$. Unless the equatorial rotational velocities are significantly larger than the $v\,\sin i$ values, this finding is probably in line with theoretical expectations. NGC\,2004/B30, a star of uncertain evolutionary status located in the Blue Hertzsprung Gap, clearly shows signs of mixing in its atmosphere. To verify the effects due to rotational mixing will therefore require homogeneous analysis of statistically significant samples of low-metallicity main-sequence B stars over a wide range of rotational velocities.
\end{abstract}





\section{INTRODUCTION}
The study of chemical abundances in evolved and main-sequence stars in
nearby galaxies is nowadays possible with 8m-class telescopes. Stellar
abundances derived from high-resolution spectra are being obtained for
different systems with interesting new results concerning their chemical evolution history.
This growing collection of stellar abundance data for other galaxies will
certainly have an impact on our understanding of
chemical evolution in different stellar populations
from environments that have undergone different histories
of star formation and heavy-element enrichment.
As nearby satellite galaxies with lower overall metallicities than the Milky Way, the Magellanic Clouds (MCs)
are prime targets for such high-resolution studies.

For the massive early-type stellar population in the Large Magellanic Cloud (LMC),
Korn et al. (2002, hereafter Paper~I) presented non-LTE abundances for four main-sequence targets
of spectral type B in the young populous cluster NGC\,2004. This was the first study
to probe the CNO content of unevolved LMC stars. The stellar CNO abundances were found to be
in excellent agreement with the LMC H\,{\sc ii}-region abundances. In particular, Paper~I confirmed
the low nitrogen abundance obtained previously from H\,{\sc ii} regions,
which is much lower than that obtained for the Milky Way.
\citet{smith02} studied a sample of M giants in the LMC and showed that
the [O/Fe] versus [Fe/H] relation obtained for these lower-mass
red giants, along with the B-star results, indicates consistently that [O/Fe] in the LMC is low when compared
with the Milky Way trend. This finding is compatible with models of chemical evolution that allow, for example, for a higher
ratio of supernovae Type Ia to Type II \citep{pagel98}, when compared to the Milky Way.

The study of the chemical content of the pristine gas in the MCs based on
spectroscopy of young early-type stars, however, must take into consideration the fact that these
stars rotate rapidly: \citet{keller04} gives a peak $v$\,$\sin$\,$i$ of around 200\,km\,s$^{-1}$ for NGC\,2004 measured from 50 main-sequence B stars. Rotation has an impact on the observed abundances as well as on a star's
evolution. Paper~I thus concentrated on stars with low $v$\,$\sin$\,$i$ (which does not necessarily indicate that these targets are
slow rotators), as these produce, in principle, more reliable
abundances obtained from unblended features.

Studies of rotation and mixing in early type stars, however, call for
larger samples of stars; probing the effects of rotation and
mixing on the main sequence, as well as in evolved stars, is crucial in order
to constrain the models. One goal of this paper is thus to complement the sample analysed in Paper~I with stars having
higher $v$\,$\sin$\,$i$ and investigate the results of mixing on the main sequence.
Moreover, the study of rotation and mixing in MC stars becomes
even more interesting as these dwarf irregular galaxies have an overall
lower metallicity than the Milky Way, and thus providing environments for studying the effects of mixing and rotation that differ fundamentally from the Galaxy.

CNO abundances are expected to be altered by rotation-induced. As an example, \citet{maeder01} predict a 9\,M$_\odot$ star of Small Magellanic Cloud (SMC) metallicity (Z\,=\,0.004) with an initial rotational velocity of $v$\,=\,300\,km\,s$^{-1}$ will show a log(N/O) ratio 0.4\,dex{\footnote{As discussed by \cite{lennon03}, this value is based on solar abundance {\sl ratios} as baseline abundances, an inappropriate assumption for the MCs. The actual N enhancement is thus expected to be larger.} above the initial value at the end of its main-sequence lifetime. Coupled with some depletion of carbon, this should produce a detectable mixing signature, {\sl if\/} the inclination angle $i$ is favourable to allow for a spectroscopic analysis. Indirect evidence for effects of mixing come from stars whose progenitors are the stars we study in this paper. For SMC A-type supergiants, \citet{venn99} reports a wide range of nitrogen enhancements with respect to the baseline value from H\,{\sc ii} regions which cannot be explained by the first dredge-up alone. Instead, this range finds its natural explanation in partial CNO mixing on the main sequence coupled with the results of the first dredge-up.
To test whether mixing can be measured directly, we extend in this paper previous abundance analysis work on B stars in NGC\,2004 to higher $v$\,$\sin$\,$i$ values.

\section{UVES OBSERVATIONS}
The observations of the targets C8 and B18 are described in detail in \citet{korn02}. Here, we only give a brief overview of the data quality. The slit width was set to 1.2$\arcsec$ to yield a resolving power of $R$\,=\,40\,000. Subsequent rebinning by a factor of 2 results in an effective resolution of 20\,000. Depending on the apparent magnitude of the target, we took one (star C8) or two (B18 and B24) exposures of 1\,hr each to reach a signal-to-noise ratio (S/N) of 100 per pixel or more for the rebinned spectra.

One additional target, B24, is not mentioned in Paper~I. According to \citet{balona93}, it is an eclipsing binary with orbital period $P$\,=\,3.85\,days. We have two 1\,hr exposures taken nearly 1 month apart (12th of December, 2000 and 9th of January, 2001). Radial velocity variations are obvious from these two spectra, but we find no spectroscopic evidence for binarity. We therefore decided to analyse it as if it were a single star. We note that this is the only target for which we can directly infer something about the actual rotational velocity $v$, as the inclination angle $i$ has to be close to 90$\degr$.
\begin{figure*}
\begin{center}
\hspace*{-2cm}
\includegraphics[angle=-90,scale=0.6]{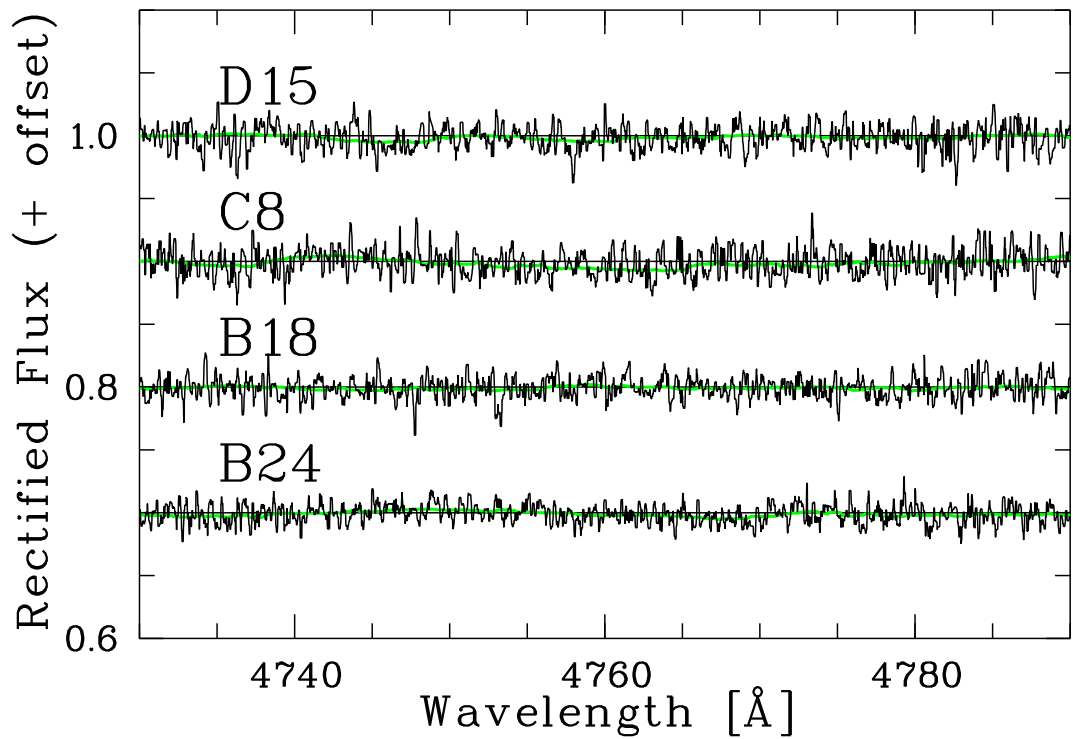}
\end{center}
\caption{Wavelength region of echelle order 98 in the slow-rotator D15 and the three program stars, offset with respect to one another for clarity. The interactively normalized, radial-velocity corrected and rebinned ($R$\,=\,20\,000) spectra are shown. The thick [green] lines are the same spectra median-filtered with a width of 200 pixels to show the residual run of the continuum. These residuals are mostly confined to $\pm$\,0.5\,\%. [See the electronic edition of the Journal for a colour version of this figure.]}
\label{fig0}
\end{figure*}

We make use of the UVES pipeline output. For the normalization, we use low-order polynomials defined by points set in continuum windows as defined by the slow rotator NGC2004/D15 from the same observing run. In B stars, such windows exist even at high $v$\,$\sin$\,$i$ values. As D15 has stellar parameters very similar to those of the program stars and is expected to have the same composition, this interactive procedure produces a very reliable first normalization.

As an example of this procedure, Figure~1 shows the wavelength region of echelle order 98 (free spectra range 4735\,--\,4783\,\AA) in D15 and the three program stars. Even though LTE synthesis tools (such as {\sc synspec}, cf. http://www.lsw.uni-heidelberg.de/cgi-bin/websynspec.cgi) predict some C\,{\sc ii} and N\,{\sc ii} lines in this spectral region, none of these are seen in D15. As for carbon, this is due to the fact that the C\,{\sc ii} features are much weaker under the more realistic non-LTE assumption. The nitrogen features are much weaker than expected due to the extraordinary low baseline nitrogen abundance of the LMC. This 60\,\AA\ stretch of effectively line-free spectra thus allows us to evaluate the inner-order continuum shape and determine how successful the normalization procedure using D15 is.

As can be appreciated from inspecting Figure~1, the interactive normalization is a rather good first approximation. Median-filtering the spectra with a width of 200 pixels reveals the continuum residuals which have three main features. They are broad (typically 10\,\AA), vary rather slowly and are mostly confined to $\pm$\,0.5\,\% of the normalized flux. This normalization is subsequently refined locally (see Section \ref{specana}). We are thus confident that we can successfully extract quantitative abundance information from spectral features as shallow as 2\,\% central line depth.

\section{ANALYSIS}
The analysis of the spectra is based on line-blanketed LTE model atmospheres computed with the ATLAS9 program. It produces plane-parallel homogeneous and static atmospheres \citep{kurucz93}. For our calculations we settled for the following parameters: 15 iterations, 64 depth levels, frequency grid LITTLE (1212 points), $\xi_{\rm atmo}= 2$\,$\mathrm{km\,s}^{-1}$. For this sample of LMC stars, we considered a metallicity of [$m$/H]\,=\,$-0.5$ (Paper~I). We performed a purely spectroscopic analysis to determine the effective temperature $T_{\rm eff}$, the surface gravity $\log$\,$g$, the microturbulence $\xi$ and the abundances of the elements C, N, O, Mg, Al and Si.

\subsection{Non-LTE Line Formation}
For a more realistic treatment of line formation in early-type stellar atmospheres, we use a statistical equilibrium approach (non-LTE) throughout. The code DETAIL (Giddings 1981, K. Butler 1996, priv.\,comm.) was used to calculate level populations and SURFACE (Becker 1997, priv.\,comm.) was used to compute the line profiles. The following model atoms were used: C \citep{przybillaC01}, N \citep{przybillaN01}, O \citep{becker88}, Mg \citep{przybillaMg01}, Al \citep{dufton86} and Si \citep{becker90}.

\subsection{Spectroscopic Analysis}\label{specana}
Due to the fact that the blending of spectral lines becomes more severe as $v$\,$\sin i$ increases, the analysis of rapidly rotating stars is less straightforward than for stars with low $\sin i$.

\begin{figure*}
\begin{center}
\includegraphics[angle=0,scale=0.35]{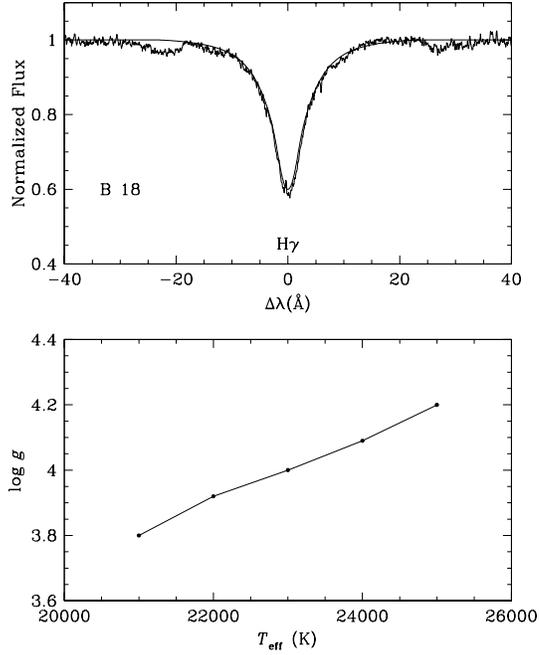}
\end{center}
\vspace*{-0.5cm}
\caption{Example of how the H$\gamma$ profile constrains the stellar parameters. The main dependencies are $T_{\rm eff}$ and $\log$\,$g$, [$m$/Fe] and $v$\,$\sin$\,$i$ play a minor role (not shown here).}
\label{fig1}
\end{figure*}

The spectral lines are generally blended and the abundance analysis becomes more difficult due to three important effects:
(1) the continuum is lowered and the difficulty of tracing it increases with the number of blended lines; (2) in blends of lines of the same element we have to find the best simultaneous fit to all components that we see as one profile; and (3) in blends caused by a mixture of lines of different elements we have to consider all components fixing the abundance of one element by isolated lines before fitting the second/third element. These effects make the integration of single-line equivalent widths practically impossible. Below we describe how we adjusted our codes to deal with high $v$\,$\sin i$ stars and blended lines.

\begin{figure*}
\begin{center}
\includegraphics[angle=-90,scale=0.29]{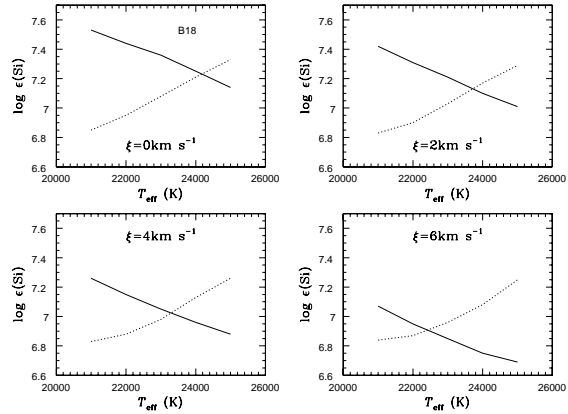}
\end{center}
\vspace*{1.6cm}
\caption{Example of how the microturbulence $\xi$ influences the derivation of $T_{\rm eff}$. Solid lines are fitting curves for Si\,{\sc iii}, dotted lines are those for Si\,{\sc ii}. In conjunction with O\,{\sc ii} lines a final $\xi$ is chosen that sets $T_{\rm eff}$ and in turn $\log$\,$g$.}
\label{fig2}
\end{figure*}
In order to synthesize the blended lines we first calculate the populations of each element individually using DETAIL and then compute the line profile with SURFACE fixing the non-LTE population of one of the components to reproduce the blend observed. Finally, before the comparison with the stellar spectrum, we convolve the synthetic profiles with a standard rotational broadening function, using $v\,\sin i$ as a free parameter. In the spectral synthesis, instrumental profile and limb darkening (with coefficients interpolated from the tables of \cite{wade85}) are also considered. For each spectral feature, the continuum is determined locally and adjusted in each iteration step, if needed.

We determine $T_{\rm eff}$ from the ionization equilibrium of Si\,{\sc ii/iii}. Based on a grid of ATLAS9 model atmospheres in the ranges 20\,000\,K $\le$ $T_{\rm eff}$ $\le$ 26\,000 K and 3.2 $\le$ $\log$\,$g$ $\le$ 4.0, the line profiles of the Balmer line $H\gamma$ were fitted to observed spectra for five values of $T_{\rm eff}$, obtaining a line in the ($T_{\rm eff}$--$\log$\,$g$)-plane (Fig.~\ref{fig1}). Along this curve in the ($T_{\rm eff}$--$\log$\,$g$)-plane, five parameter pairs ($T_{\rm eff}$, $\log$\,$g$) were selected to compute model atmospheres. Following this, we derived silicon abundances for lines of the two ionization stages (Si\,{\sc ii} and {\sc iii}) for each model atmosphere. The intersection point in the $\log$\,$\varepsilon$(Si)-$T_{\rm eff}$ diagram yields the effective temperature and the gravity. The solution for $T_{\rm eff}$ depends somewhat on the assumed value of $\xi$ (Fig. \ref{fig2}) requiring an iterative process using silicon and oxygen lines to find the best solution.

\begin{table*}
\caption{Chemical analysis of the slow rotator NGC\,2004/D15 based on the stellar parameters given in Paper~I: $T_{\rm eff}$\,=\,22500\,K, $\log$\,$g$\,=\,3.80, [$m$/Fe]\,=\,$-0.5$, $\xi$\,=\,0\,km\,s$^{-1}$, $v$\,$\sin$\,$i$\,=\,45\,km\,s$^{-1}$. The abundances agree very well.}
\label{table1}
\begin{center}
\begin{tabular}{ccccccc}
\hline
\noalign{\smallskip}
NGC\,2004/D15 & $\log$\,$\varepsilon$(C) & $\log$\,$\varepsilon$(N) & $\log$\,$\varepsilon$(O) & $\log$\,$\varepsilon$(Mg) & $\log$\,$\varepsilon$(Si) \\
\noalign{\smallskip}
\hline
\noalign{\smallskip}
Korn et al.~(2002) & 8.04 & 6.95 & 8.29 & 7.35 & 7.04 \\
$\pm$ & 0.2 & 0.2 & 0.2 & 0.2 & 0.2 \\
\noalign{\smallskip}
\hline
\noalign{\smallskip}
this work & 8.00 & 6.99 & 8.29 & 7.44 & 6.98 \\
$\pm$ & 0.2 & 0.2 & 0.2 & 0.2 & 0.2 \\
\noalign{\smallskip}
\hline
\end{tabular}
\begin{list}{}{}
\item
$\log \varepsilon \mathrm{(X)}\equiv\log n_\mathrm{X}/n_\mathrm{H} + 12$
\end{list}
\end{center}
\end{table*}
\begin{figure*}
\begin{center}
\includegraphics[angle=0,scale=0.65]{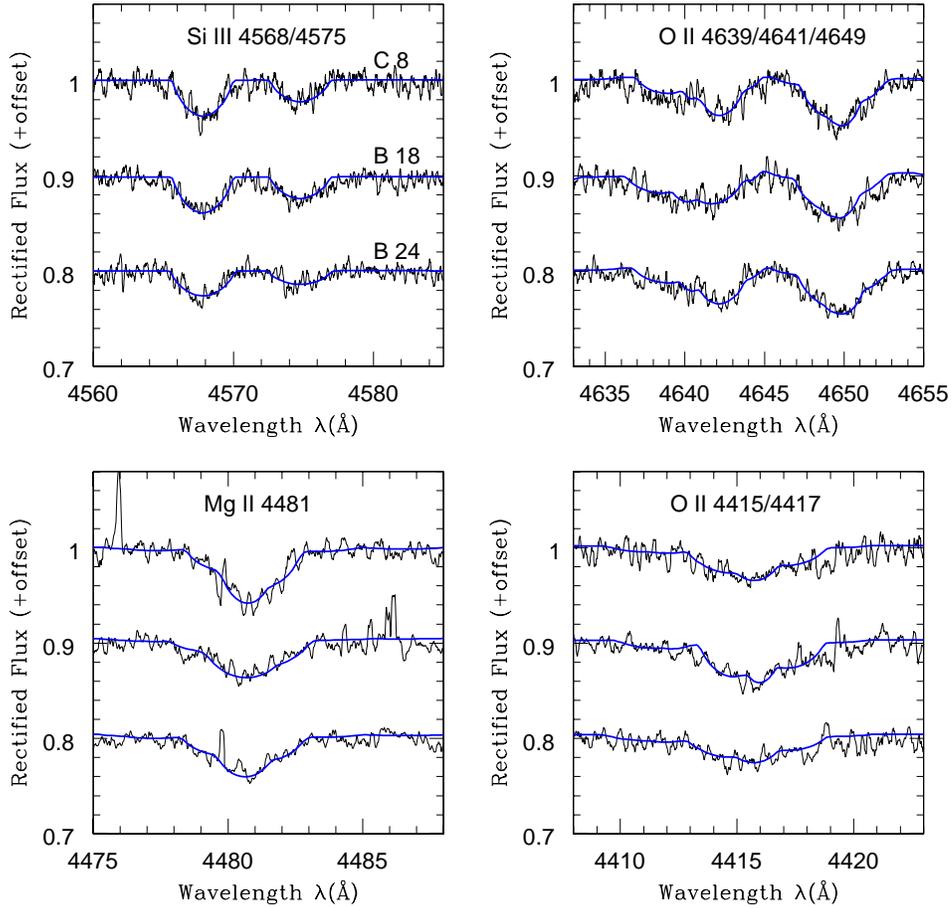}
\end{center}
\caption{Final profile fits for some of the lines used in the spectroscopic abundance analysis shown for selected wavelength regions for all target stars. [See the electronic edition of the Journal for a colour version of this figure.]}
\label{fig4}
\end{figure*}
For each solution of $T_{\rm eff}$ from Fig.~\ref{fig2}, we analyse O\,{\sc ii} and Si lines to find a simultaneous solution for both elements. This procedure singles out a final $T_{\rm eff}$--$\log$\,$g$ pair which fulfills the ionization equilibrium of silicon and the microturbulence constraint for silicon and oxygen. In order to find the best overall fit, we let the values of $v$\,$\sin$\,$i$, $\xi$ and the chemical abundances vary and minimize $\chi^{2}$ with respect to the observation.

We thus simultaneously derive $T_{\rm eff}$, $\log$\,$g$, $\xi$, $v$\,$\sin$\,$i$, $\log$\,$\varepsilon$(Si) and $\log$\,$\varepsilon$(O). {bf As a final check of the stellar parameters, the H$\beta$ profile is inspected. We find good agreement with H$\gamma$.} Afterwards, the abundances of the other elements are calculated with some allowance for $v$\,$\sin$\,$i$ to vary.

As a consistency check on the methodology adopted in this study for the high $v$\,$\sin i$ stars in NGC\,2004, we first analysed one sharp-lined star from the same cluster, D15, previously studied in Paper~I. The results from this comparison are summarized in Table~1. There is excellent agreement, considering the uncertainties, between the two sets of solutions in terms of stellar parameters (not shown) and elemental abundances.

For this star, we also experimented with photometric indicators of $T_{\rm eff}$, notably the reddening-free $Q$-parameter calibration by \citet{daflon99}. The $Q$ value for this star is at the lower limit of validity for the calibration curve where $T_{\rm eff}$ ceases to be sensitive to $Q$. We found a markedly lower effective temperature ($T_{\rm eff}$\,=\,19\,500\,K) which also entailed a lower gravity ($\log$\,$g$\,=\,3.5) via the use of H$\gamma$. These modified stellar parameters clearly leave their imprint on the chemical abundances: the magnesium abundance decreases by 0.4\,dex, while oxygen increases by 0.25\,dex. These modified abundances do not seem plausible. For consistency with Paper~I, we refrained from exploring this alternative temperature scale and its implications further and performed all analyses fully spectroscopically.

Below, we comment on each element and the respective lines found in the spectra of the program stars.\\
{\sl Carbon}: C\,{\sc ii} 4267 is the only line that we could identify in the spectra of the sample stars. It is strong and unblended and, in contrast what was found in  earlier works \citep{rolleston03}, seems to return sensible absolute abundances in non-LTE.\\
{\sl Nitrogen}: We could identify two isolated lines (N\,{\sc ii} 3995 and 4630) in the three stars, plus one blended line (N\,{\sc ii} 4643) in the case of B24.\\
{\sl Oxygen}: This is the most reliable element to be analysed because it has more lines in this part of the spectrum than any other elements. We analysed a variety of O\,{\sc ii} lines, both isolated and blended. The adopted linelist can be found in \cite{daflon01}.\\
{\sl Aluminium}: The spectra show a very weak line Al\,{\sc iii} 4480 in the blue wing of the strong line Mg\,{\sc ii} 4481 which results in a large uncertainty for $\log$\,$\varepsilon$(Al). The aluminium abundance in all stars seems compatible with a single value of around $\log$\,$\varepsilon$(Al)\,=\,5.7.\\
{\sl Magnesium}: We had to fix the aluminium abundance in order to reproduce the blend Al\,{\sc iii} 4480/Mg\,{\sc ii} 4481 from which we derived the Mg abundance.\\
{\sl Silicon}: The ionization equilibrium was solved for lines of Si\,{\sc ii/iii}. The Si\,{\sc iii} lines at 4553\,--\,4574\,\AA\ were all isolated, while the Si\,{\sc ii} lines at 4128/4131\,\AA\ are located in the wing of H$\delta$ and are blended by O\,{\sc ii} lines. The Si\,{\sc iv} line at 4116\,\AA\ was found to be too weak to utilize, due to both the high $v$\,$\sin$\,$i$ values and the relatively low values of $T_{\rm eff}$.

Fig.~\ref{fig4} shows some profile fits of blended lines, some strategic lines used in the derivation of the stellar parameters and some abundance indicators such as Mg\,{\sc ii} 4481.

\subsection{Abundance Uncertainties}
The sources of uncertainty in abundance determinations come from uncertainties in stellar parameters ($T_{\rm eff}$, $\log$\,$g$, $\xi$, $v$\,$\sin$\,$i$), the uncertainty in the position of the continuum, the completeness of the line list used and limitations of the model atoms and the entire spectral synthesis process.

To estimate typical error bars, we calculated the abundance uncertainties due to each para\-meter independently using O\,{\sc ii} lines and combined that value in quadrature with the continuum uncertainty. The errors in the element abundances are dominated by the uncertainties in the definition of the continuum (in many cases it is very difficult to distinguish between weak spectral features and noise; see Fig.~\ref{fig4}). We found a total uncertainty of the oxygen abundance of 0.3\,dex. We did not analyse the uncertainty of each line of the other elements due to the atmospheric parameters because the continuum placement has the largest contribution to the error budget. Since all elements are represented by lines with central line depths of 2\,\% or more, we adopt 0.3\,dex as a typical value for the total uncertainty of elemental abundances derived here. However, this uncertainty does not include systematic effects potentially arising from inappropriate modelling assumptions.

\begin{table*}
\footnotesize
\caption[]{Spectroscopic stellar parameters and final abundances for the five elements under investigation. The values for $T_{\rm eff}$, $\log$\,$g$, $\xi$, $v$\,$\sin$\,$i$ and $\log$\,$\varepsilon$(X) are rounded to the nearest 100\,K, 0.05\,dex, 1\,km\,s$^{-1}$, 10\,km\,s$^{-1}$ and 0.15\,dex, respectively. Note that the aluminium abundance was set at $\log$\,$\varepsilon$(Al)\,=\,5.7 to extract the magnesium abundance from the Al\,{\sc iii} 4480/Mg\,{\sc ii} 4481 blend.
LMC H\,{\sc ii}-region abundances are quoted from \cite{garnett99}, and the solar abundances from \cite{asplund04}}
\label{stellpar}
$$
\begin{array}{lccccccccc}
\hline
\noalign{\smallskip}
\mathrm{star} & {T_{\rm eff}} & \log\,g & \xi & v\,\sin i & \log\,\varepsilon$(C)$ & \log\,\varepsilon$(N)$ & \log\,\varepsilon$(O)$ & \log\,\varepsilon$(Mg)$ & \log\,\varepsilon$(Si)$  \\
& [$K$] &  & \mathrm{[km\,s}^{-1}] & \mathrm{[km\,s}^{-1}] \\[1mm]
\hline
\noalign{\smallskip}
\mathrm{NGC\,2004/C8} & 22700 & 3.80 & 5 & 120 & 8.05 & 6.95 & 8.40 & 7.15 & 7.10  \\
\pm         & {\it 1000}  & {\it 0.2}  &  {\it 2} & {\it 20} & {\it 0.3}  & {\it 0.3} & {\it 0.3} & {\it 0.3}   & {\it 0.3} \\
\noalign{\smallskip}
\mathrm{NGC\,2004/B18} & 23500 & 4.05 & 4 & 130 & 7.90 & 7.00 & 8.40 & 7.10 & 7.00  \\
\pm         & {\it 1000}  & {\it 0.2}  &  {\it 2} & {\it 20} & {\it 0.3}  & {\it 0.3} & {\it 0.3} & {\it 0.3}   & {\it 0.3} \\
\noalign{\smallskip}
\mathrm{NGC\,2004/B24} & 23000 & 3.70 & 0 & 130 & 8.00 & 7.10 & 8.40 & 7.20 & 7.10 \\
\pm         & {\it 1000}  & {\it 0.2}  &  {\it 2} & {\it 20} & {\it 0.3}  & {\it 0.3} & {\it 0.3} & {\it 0.3}   & {\it 0.3} \\
\noalign{\smallskip}
\hline
\noalign{\smallskip}
\mathrm{mean(this \,work)} & & & & & 7.98 & 7.02 & 8.40 & 7.15 & 7.07 \\
\noalign{\smallskip}
\mathrm{mean(Paper~I)} & & & & & 8.06 & 7.01 & 8.37 & 7.37 & 7.10 \\
\noalign{\smallskip}
\mathrm{mean(H\,II)}_{\rm LMC} & & & & & 7.90 & 6.90 & 8.40 & - & 6.70 \\
\noalign{\smallskip}
\mathrm{Sun} & & & & & 8.39 & 7.78 & 8.66 & 7.53 & 7.51 \\
\noalign{\smallskip}
\hline
\end{array}
$$
\end{table*}

\section{DISCUSSION}
Table \ref{stellpar} lists the final stellar parameters and the derived chemical abundances for the three program stars. Assuming error bars of 0.3\,dex as reasoned above, all stars have practically identical abundances. In fact, the abundances often agree to within 0.1\,dex, a very encouraging result. Good agreement was certainly expected for elements like magnesium or silicon, not necessarily so for nitrogen and carbon.
Formulated relative to the latest solar-system abundance distribution \citep{asplund04}, the mean abundances are $-$0.26\,dex in [O/H] (using the customary square-bracket notation\footnote{$\mathrm{[X/Y]\equiv \log(X/Y)_\star-\log(X/Y)_\odot}$}) to $-$0.76\,dex in [N/H] confirming the extraordinary low nitrogen content peculiar to the MCs (\cite{korn01}; Rolleston et al.\ 2003). Carbon, magnesium and silicon show [X/H] values of $-0.41$, $-0.38$ and $-0.44$\,dex respectively, typical values for young LMC objects. Even though iron lines are generally to weak to be measured in the program stars of this paper, we re-iterate from Paper~I that the present-day ratio of $\alpha$-capture elements (here, O, Mg and Si) to iron in the LMC is at best solar ([$\alpha$/Fe]\,=\,0), more likely slightly negative.

In comparing the mean abundances with values from LMC H\,{\sc ii} regions and the main-sequence B stars from Paper~I, three features are worth commenting on. The abundances for the elements C, N and O agree very well indicating that the atmospheres of the target stars are unmixed. While we have no H\,{\sc ii}-region value for Mg to compare with, the mean stellar Mg abundances of the two samples differ by 0.2\,dex. We have no explanation for this finding, Mg is certainly not expected to vary significantly on the main sequence. Finally, the stellar silicon abundance is once again found to be systematically above the nebular one. This offset is likely caused by some silicon locked up in interstellar grains.

All three stars have $v$\,$\sin$\,i values close to 130\,km\,s$^{-1}$ as measured from the strategic lines of silicon and oxygen. For an optimal fit, other features (e.g.~Mg\,{\sc ii} 4481 or C\,{\sc ii} 4267) require lower or higher values such that the numbers given in Table~\ref{stellpar} are somewhat model dependent. \citet{keller04} reports $v$\,$\sin$\,i values for all of our targets from medium-resolution spectroscopy (the same spectroscopy that was used to preselect our targets in the first place). For C8 and B24, our estimates agree very well, for B18 our value is 60\,km\,s$^{-1}$ smaller (likewise for D15 from Paper~I). This is in line with error estimates performed by \citet{keller04} and does not point toward significant systematic differences.

In Figure~\ref{fig6} the results are plotted in a variety of ways. The top panel shows the program stars in a Kiel diagram from which their evolutionary status can be estimated. In comparison with the non-rotating evolutionary tracks of \citet{schaerer93}, all three objects fall onto the main-sequence band, roughly half way between the zero-age main sequence and the turnoff. In comparison with the low-$v$\,$\sin$\,$i$ stars from Paper~I, the stellar parameters derived here are in no way peculiar. These stars thus support the cluster age estimate of (21\,$\pm$\,3)\,Gyr made in Paper~I.

The middle panel of Figure~\ref{fig6} displays the logarithmic abundance ratio of nitrogen to oxygen ($\log$\,(N/O)) as a function of $\log$\,$g$. The effects of rotational mixing would produce an increasing trend of $\log$\,(N/O) with decreasing $\log$\,$g$ among main sequence stars. Such a trend is, however, not seen from this small sample of stars. B30 from \citet{korn00} does show an enhanced $\log$\,(N/O) ratio, but as it falls into the Blue Hertzsprung Gap, its evolutionary status is uncertain: it might be in an advance stage of evolution on a blue-loop excursion (potentially explaining the lower $\log$\,$g$/higher luminosity) in which case the high $\log$\,(N/O) ratio would be caused by the first dredge-up taking place in the red-giant phase. This complication is exactly the reason why the predictions of rotational mixing must be tested on the main sequence.

\begin{figure*}
\begin{center}
\includegraphics[angle=0,scale=0.60]{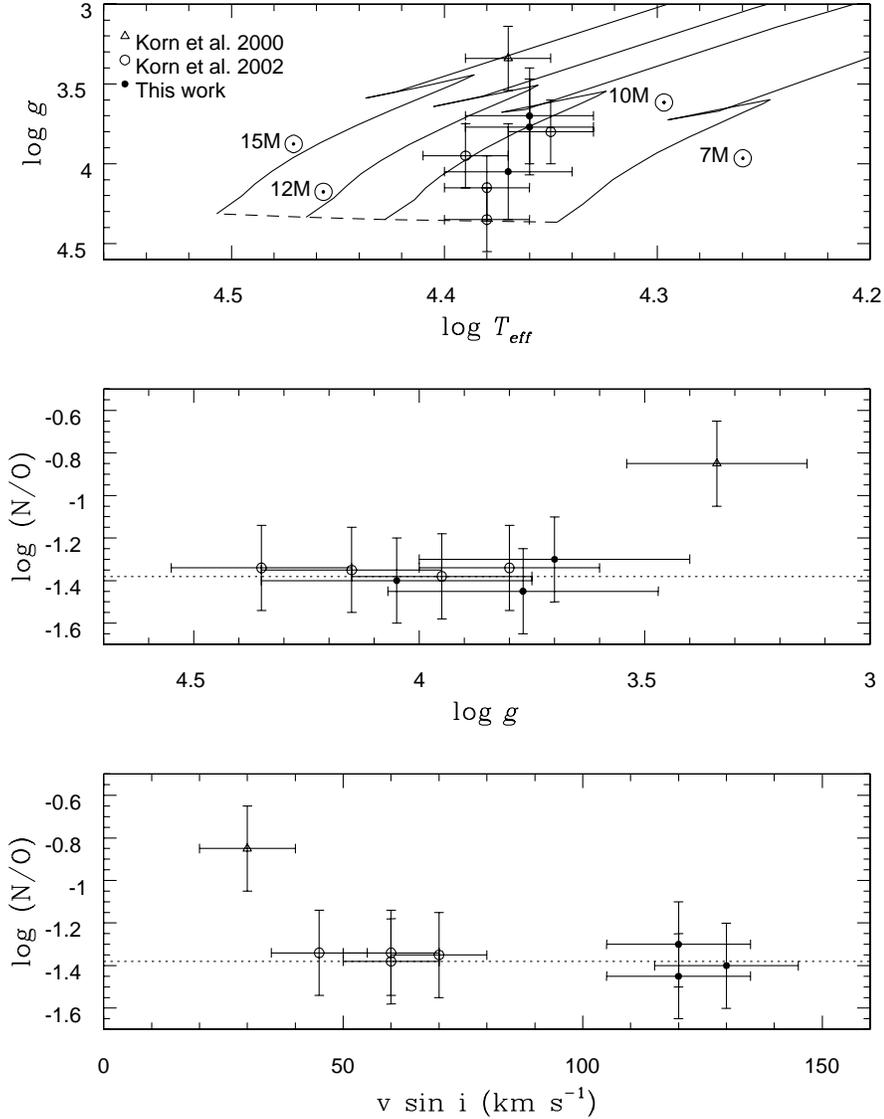}
\end{center}
\caption{{\sl Top:} Program stars (filled circles) and those of the same cluster analysed by Korn et al.~(2000, 2002; triangle and open circles, respectively) in comparison with evolutionary tracks for non-rotating stars of \citet{schaerer93}. The dashed lines indicates the zero-age main sequence for Z\,=\,0.008. Except for B30 in the Blue Hertzsprung Gap, all stars fall between the tracks for 8 and 11\,M$_\odot$ stars.
{\sl Middle:} The $\log$\,(N/O) ratio as a function of $\log$\,$g$ as an indicator of the evolutionary status of the stars. The N/O ratio is expected to increase as the stars evolve along the main sequence. No evolution of $\log$\,(N/O) is seen on the main sequence. B30 shows an enhanced N/O ratio. The dashed line indicates the mean log\,(N/O) of the three program stars.
{\sl Bottom:} The $\log$\,(N/O) ratio as a function of $v$\,$\sin$\,$i$. No evidence for enhanced N/O ratios is found among the fast rotators. B30, however, is enhanced in N and depleted in O. We note that its low $v$\,$\sin$\,$i$ value does not necessarily imply slow rotation. The dashed line indicates the mean log\,(N/O) of the three program stars.
}
\label{fig6}
\end{figure*}

Finally, the lower panel shows $\log$\,(N/O) as a function of $v$\,$\sin$\,$i$. The program stars show the same unmixed $\log$\,(N/O) ratio as the stars from Paper~I, only B30 from \citet{korn00} shows an enhanced value. We can only speculate that B30 is an initially fast-rotating star that is viewed pole-on or a star that has spun down upon becoming a giant and is now on a blue-loop excursion. As fast-rotating stars evolve to higher luminosities \citep{fliegner96}, the prior scenario could explain its seemingly high mass (cf.~top panel). If the latter scenario applies, one would have to explain the contradicting theoretical expectation that a 9\,M$_\odot$ star of SMC metallicity will spin up during the blue-loop phase \citep{maeder01}.

From the seven stars homogeneously analysed here and in Paper~I, we find no evidence for mixing of CN-cycled material into the atmospheres of main-sequence B stars in NGC\,2004. However, this does not imply that processes like rotational mixing do not exist. It only shows that, given the selection biases, we have not looked at large enough samples of stars. The fact that we are only able to analyse stars with $v$\,$\sin$\,$i$ up to 150\,km\,s$^{-1}$ is likely enough of a bias to explain the absence of direct observational support for rotational mixing (cf.~\citet{fliegner96}). The example given in {\S}1 clearly overestimates the signature expected for the targets of this study in three ways: an SMC metallicity was adopted, the initial equatorial rotation rate was assumed to be 300\,km\,s$^{-1}$ and the log(N/O) ratio was calculated at the end of the main-sequence evolution. For a 9\,M$_\odot$ LMC star half way across the main-sequence band and currently rotating at $v$\,$\approx$\,130\,km\,s$^{-1}$ (like B24), the log(N/O) ratio is expected to be far less enhanced (G.~Meynet, priv.\,comm.).

One must not, however, forget that there are individual main-sequence stars like NGC\,1818/D1 from Paper~ I whose abundance signature is highly indicative of some mixing process taking place. Whether this process is rotational mixing or binary mass transfer, is difficult to tell for a single star. This highlights the need to analyse statistically significant samples of main-sequence B stars.

\section{CONCLUSIONS}
We have modified existing codes to be able to analyse B stars rotating up to $v$\,$\sin$\,$i$\,$\approx$\,150\,km\,s$^{-1}$. As the spectra of B stars are generally rather clean, blending of diagnostic features for the stellar parameter and chemical abundance determination can be dealt with rather reliably. The large $v$\,$\sin$\,$i$ values mainly result in larger error bars and some elements becoming undetectable (e.g. iron). Systematic errors can be controlled, as long as the data quality (in terms of S/N, pixel-to-pixel variations and echelle-order blaze residuals) is high.

Among the three fast rotators, we find no evidence for altered chemical abundances, i.e., no enhanced nitrogen or depleted carbon as would be expected from contamination with CN-processed material brought up by rotational mixing. Even though the number of stars analysed is small, this puts some constraints on the predictions from rotating stellar evolution models. It is, however, difficult to quantify these constraints as we do not know the true rotation rates of the program stars (with the notable exception of the eclipsing binary B24). Statistically much more significant samples of main-sequence B stars in MC clusters are currently being analysed (the FLAMES survey of massive stars in the Magellanic Clouds at the VLT, see \cite{evans05} for an overview). These data sets will likely be able to answer the question of to what extent observations support the theoretical predictions of rotational mixing.

Judging from this work, it will be very important to push the limit of $v$\,$\sin$\,$i$ to values as high as possible. It is likely that effects only become measurable above $v$\,$\sin$\,$i$\,$\approx$\,150\,km\,s$^{-1}$ (cf.~the example given in {\S}1) or at metallicities lower than those found in the present-day LMC. Analysing metal-poor stars rotating at such high rates will hold a variety of challenges to be tackled in future works. Such efforts must be undertaken to reliably model, e.g., the chemical evolution of nitrogen and the signatures of the first supernovae encoded in the spectra of the most metal-poor stars found in the Galactic halo.

\begin{acknowledgements}
Support by the Leopoldina Foundation/Germany under grant
BMBF-LPD 9901/8-87 (A.\,J.\,K.), from a CNPq fellowship (M.\,F.\,N.) and an AURA Inc.\ US Gemini fellowship (K.\,C.) is gratefully acknowledged.
\end{acknowledgements}

\end{document}